\documentclass[letter]{IEEEtran}
\usepackage{latexsym,,amssymb,amsmath,graphicx,epsf,cite,bbm,float,subfig}
\usepackage{ifpdf}
\usepackage{epstopdf}

\def\ninept{\def\baselinestretch{0.99}}
\ninept

\newcommand{\be}{\begin{equation}}
\newcommand{\ee}{\end{equation}}
\newcommand{\bea}{\begin{eqnarray}}
\newcommand{\eea}{\end{eqnarray}}
\newcommand{\nn}{\nonumber}

\newcommand{\MB}{\left[\begin{array}}
\newcommand{\ME}{\end{array}\right]}

\renewcommand{\vec}[1]{\mbox{\boldmath${#1}$}}

\newcommand{\vu}{\vec{u}}
\newcommand{\vw}{\vec{w}}
\newcommand{\mC}{\vec{C}}
\newcommand{\va}{\vec{a}}
\newcommand{\vz}{\vec{z}}

\newcommand{\ei}{\end{itemize}}
\newcommand{\bi}{\begin{itemize}}

\newcommand{\vc}{\mbox{$\vec{c}$}}
\newcommand{\vphi}{\mbox{$\vec{\phi}$}}

\newcommand{\calN}{{\cal N}}
\newcommand{\defi}{\stackrel{\bigtriangleup}{=}}
\newcommand{\eps}{\mbox{$\epsilon$}}

\AtBeginDocument{
\addtolength{\abovedisplayskip}{-0.4ex}
\addtolength{\abovedisplayshortskip}{-0.4ex}
\addtolength{\belowdisplayskip}{-0.4ex}
\addtolength{\belowdisplayshortskip}{-0.4ex}
\addtolength{\belowcaptionskip}{-3.6ex}
}

\begin{document}

\title{Single Bit and Reduced Dimension Diffusion Strategies Over Distributed~Networks} 
\author{Muhammed O. Sayin*, Suleyman S. Kozat,~{\em Senior Member, IEEE}
\thanks{Muhammed O. Sayin (m\_sayin@ug.bilkent.edu.tr) and Suleyman S. 
  Kozat (kozat@ee.bilkent.edu.tr) are with the Department of Electrical and 
  Electronics Engineering, Bilkent University, Ankara, Turkey. }}

\maketitle
\begin{abstract}
We introduce novel diffusion based adaptive estimation strategies for
distributed networks that have significantly less communication load
and achieve comparable performance to the full information exchange
configurations. After local estimates of the desired data is produced
in each node, a single bit of information (or a reduced dimensional
data vector) is generated using certain random projections of the
local estimates. This newly generated data is diffused and then used in
neighboring nodes to recover the original full information. We provide
the complete state-space description and the mean stability analysis
of our algorithms.
\end{abstract}
\vspace{-0.1in}
\section{Introduction}
\IEEEPARstart{D}{istributed} adaptive estimation utilizes a network of
nodes that observe a monitored phenomena with different view
points. This broadened perspective can be used to enhance estimation
performance or eliminate obstructions in the environment, which may
not be achieved using a single node~\cite{lopes2008}. The distributed
algorithms usually target to reach the best estimate that could be
produced when the individual nodes have access to all observations
across the whole network. However, there is naturally a trade-off
between the amount of cooperation and required communication 
among the nodes \cite{lopes2008,cattivelli2010}.

The diffusion based distributed algorithms define a strategy in which
the nodes from a predefined neighborhood could share information with
each other~\cite{lopes2008,cattivelli2010}. Such approaches are stable
against time-varying statistical profiles~\cite{lopes2008}, however,
require a high amount of communication resources. For example, in a
network of \(N\) nodes, where \(\overline{n}\) denotes the average
number of nodes in a neighborhood, then \(N\times \overline{n}\)
number of parameter estimates should be communicated among nodes on
the average at each time.

In this letter, we propose diffusion based cooperation strategies that
have significantly less communication load (e.g., a single bit of
information exchange) and achieve comparable performance to the full
information exchange configurations under certain settings. In this
framework, after local estimates of the desired vector is produced in
each node, a single bit of information (or a reduced dimensional data
vector) is generated using certain random projections of the local
estimates. This new information is diffused and used in neighboring
nodes instead of the original estimates; significantly reducing the
communication load in the network.  We only require synchronization of
this randomized projection operation, which can be achieved using
simple pilot signals \cite{sayed_book}.  Note that our approach
differs from quantization based diffusion strategies such as
\cite{xie2013} in terms of the compression of the diffused
information. In~\cite{xie2013}, a quantized parameter estimate is
exchanged among nodes to avoid infinite precision in the
transmission. Here, we substantially compress the exchanged
information, even to a single bit, and perform local adaptive
operations at each node to recover the full information vector.  In
this sense, our method is more akin to compressive sensing rather than
to a quantization framework.

Our main contributions include: 1) We propose algorithms to
significantly reduce the amount of communication between nodes for
diffusion based distributed strategies; 2) We analyze the stability of
the algorithms in the mean under certain statistical conditions; 3) We
illustrate the comparable convergence performance of these algorithms
in different numerical examples. We emphasize that although we only
provide the mean stability analysis due to space limitations, the
mean-square convergence (and tracking) analysis can be readily carried
out in a similar fashion (following \cite{lopes2008}) since we provide
the complete state-space representation.

The letter is organized as follows. In Section II, we introduce the
framework and the studied problem. The new approaches are derived in
Section III. In Section IV, we analyze the mean stability of our
approaches. Numerical examples and concluding remarks are provided in
Section V.
\vspace{-0.1in}
\section{Problem Description}
Consider the widely studied spatially distributed
framework~\cite{lopes2008,cattivelli2010}. Here, we have $N$ number of
nodes where two nodes are considered neighbors if they can exchange
information. For a node $i$, the set of indexes of its neighbors
including the index of itself is denoted by $\calN_i$. At each node,
an unknown desired vector\footnote{Although, we assume a time
  invariant desired vector, our derivations can be readily extended to
  certain non-stationary models \cite{sayed_book}.}, $\vw_o \in
\mathbbm{R}^m$, is observed through a linear model $d_i(t) = \vw_o^T
\vu_i(t) + v_i(t)$,\footnote{We represent vectors (matrices) by bold
  lower (upper) case letters. For a matrix $\vec{A}$ (or a vector
  $\vec{a}$), $\vec{A}^T$ is the transpose. $\|\vec{a}\|$ is the
  Euclidean norm. For notational simplicity we work with real data and
  all random variables have zero mean. The sign of $a$ is denoted by
  $\mathrm{sign}(a)$ ($0$ is considered positive without loss of
  generality). For a vector $\va$, $\mathrm{dim}(\va)$ denotes the
  length. The expectation of a vector or a matrix is denoted with an
  over-line, i.e. $E[\va] = \overline{\va}$. The
  $\mathrm{diag}(\vec{A})$ returns a new matrix with only the main
  diagonal of $\vec{A}$ while $\mathrm{diag}(\va)$ puts $\va$ on the
  main diagonal of the new matrix.}  assuming the observation noise is
temporally and spatially white (or independent), i.e., $v_i(t)v_j(l) =
\sigma_i^2 \delta(i-j) \delta(t-l)$, where $\delta(\cdot)$ is the
Kronecker delta and $\sigma_i^2$ is the variance of the noise. The
regression vectors are also assumed to be spatially and temporally
uncorrelated with each other and with the observation noise.  At each
node an adaptive estimation algorithm is working such as the LMS
algorithm \cite{sayed_book} given as
\begin{align*}
\vec{\phi}_i(t+1) &= (\vec{I} - \mu_i \vu_i(t) \vu_i^T(t)) \vw_i(t) + \mu_i d_i(t) \vu_i(t),
\end{align*}
$\mu_i > 0$.  As the diffusion strategy, we use the adapt-then-combine
(ATC) diffusion strategy as an example since it is shown to outperform
the combine-then-adapt diffusion and consensus strategies under
certain conditions \cite{tu}. However, our derivations also cover
these distributed strategies. In the ATC strategy, at each node $i$,
the final estimate is constructed as
\begin{align*}
\vw_i(t+1) =  \sum_{k \in \calN_i} \lambda_{i,k} \vec{\phi}_k(t+1),
\end{align*}
where $\lambda_{i,k}$'s are the combination weights $\sum_{k \in
  \calN_i} \lambda_{i,k} = 1$ and $\lambda_{i,k} \geq 0$. The
combination weights $\lambda_{i,k}$ can also be adapted in time,
affinely constrained or unconstrained \cite{kozat}. We stick to
constant-in-time weights with the simplex constraint since the
stabilization effect of such weights is demonstrated in
\cite{lopes2008}.

In the diffusion based distributed networks, whole parameter estimates
are exchanged within the neighborhood.  In the next section, we
introduce two different approaches in order to reduce the amount of
information exchange between nodes. 
\section{New Diffusion Strategies} 
\subsection{Reduced Dimension Diffusion} 
In the first approach, each node calculates a reduced dimensional
vector through a linear transformation $\vz_k(t+1) = \vec{C}(t+1)
\vphi_k(t+1)$, where $\mathrm{dim}(\vz_k(t+1))
\ll\mathrm{dim}(\vphi_k(t+1))$, and transmits $\vz_k(t+1)$ instead of
$\vphi_k(t+1)$. We use a randomized linear transformation matrix
$\vec{C}(t+1)$ where the size of the matrix determines the compression
amount. Each neighboring node uses the same $\vec{C}(t+1)$. After
receiving $\vz_k(t+1)$, a neighbor node $i$ constructs an estimate
$\va_{k}(t+1)$ of the original $\vphi_k(t+1)$ using a \emph{minimum
  disturbance criteria} \cite{sayed_book} as
\begin{align}
& \va_{k}(t+1) = \arg \min_{\va} \|\va-\va_{k}(t)\| \label{eq:minimize} \\
& \mbox{such that } \; \mC(t+1)\va=\vz_{k}(t+1), \nn
\end{align}
where  $\va_k(t) \in \mathbbm{R}^m$. Note that \eqref{eq:minimize}
yields the NLMS algorithm \cite{sayed_book} as
\begin{align}
& \va_{k}(t+1) = \va_{k}(t) + \sigma_{k} \mC(t+1)^T \big[\mC(t+1)
    \times \nn \\ & \mC(t+1)^T\big]^{-1}\big(\vz_{k}(t+1) -
  \mC(t+1)\va_{k}(t)\big), \label{eq:inv}\\ & = \left(\vec{I} -
  \sigma_{k} \vec{P}_{\mC(t+1)}\right) \va_{k}(t) + \sigma_{k}
  \vec{P}_{\mC(t+1)} \vphi_{k}(t+1), \nn
\end{align}
where a learning rate $\sigma_{k} > 0$ is also incorporated after
\eqref{eq:minimize}, $\vec{P}_{\mC(t+1)}$ is the projection matrix of
the row space of $\mC(t+1)$ or $\vec{P}_{\mC(t+1)}=\mC(t+1)^T\left[\mC(t+1)\mC(t+1)^T\right]^{-1}\mC(t+1)$
if $\mC(t+1)$
has full row rank.  

After $\va_{k}(t+1)$'s are calculated, we construct the final estimate at node $i$ as
\begin{align}
\vec{w}_i(t+1) = \lambda_{i,i} \vec{\phi}_i(t+1) + \sum_{k \in \calN_i \setminus i} \lambda_{i,k} \va_{k}(t+1).\label{comb1}
\end{align}

\noindent
{\bf Remark 1:} For a time invariant projection matrix,
$\vec{C}(t)=\vec{C}$, the exchanged estimate $\va_{k}(t)$ converges to
the projection of the original parameter estimate $\vphi_k(t)$ onto
the column space of the matrix $\vec{C}$ (provided that adaptation is
fast enough). In order to avoid biased convergence, we choose
randomized projection matrices that span the whole parameter
space.\\ {\bf Remark 2:} One can also use an ordinary LMS update to
train $\va_{k}(t)$ to avoid the inversion operation in \eqref{eq:inv},
considering $\vz_k(t+1)$ as the desired data and $\vec{C}(t+1)$ as the
regression matrix. However, since the dimensions of $\vz_{k}(\cdot)$'s
are much smaller than the dimension of $\vw_o$, e.g., in our
simulations we use scalar $z_k(\cdot)$'s with $m=1$, one can use the
NLMS update for $\va_{k}(\cdot)$'s without significant computational
increase.

In the following, we further reduce the amount of transmitted information by diffusing a single bit of information instead of a scalar.
\vspace{-0.1in}
\subsection{Single Bit Diffusion}
In this approach, we exchange only the sign of the linear transformation \(z_k(t+1) = \vc(t+1)^T\vphi_k(t+1)\). According to the transmitted sign, the neighboring node $i$ can construct an estimate \(\va_{k}(t+1)\) of \(\vphi_k(t+1)\) as 
\begin{align}
&\va_{k}(t+1) = \arg \min \|\va-\va_{k}(t)\| \label{eq:sign1} \\
&\mbox{such that} \nn\\
&\mathrm{sign}\left( \vec{c}(t+1)^T\va \right)=  \mathrm{sign} \left(z_k(t+1)\right) \mbox{and} \label{constraint}\\
&\|\va\| =1. \label{normalization}
\end{align}
\begin{figure}[t!]
\centering
\includegraphics[width=2.5in]{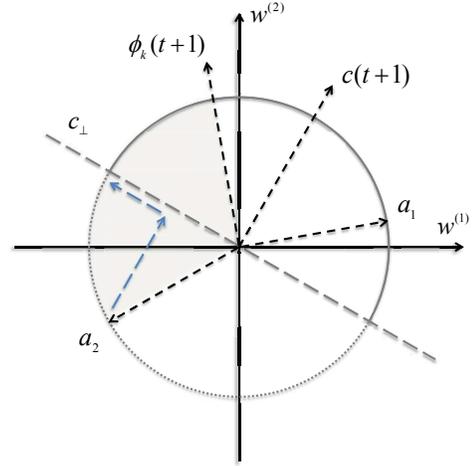}
\caption{Single bit diffusion in two dimensions, i.e., $\vec{w}_o \in \mathbbm{R}^2$. As an example, one can have  $\vec{a}_{k}(t)=\vec{a}_1$ or $\vec{a}_{k}(t)=\vec{a}_2$. $\vc_{\perp}$ denotes the vector space perpendicular to $\vec{c}(t+1)$ in two  dimensions and the shaded area represents the update region for $\va_{k}(t) = \va_2$}
\label{projection}
\end{figure}
To solve \eqref{eq:sign1}, we observe from Fig.~\ref{projection} that
we can only have two different cases for $\va_{k}(t)$. In the first
case, we have $\mathrm{sign} \left( \vec{c}(t+1)^T\va_{k}(t) \right) =
\mathrm{sign} \left(z_k(t+1)\right) $, e.g., $\va_{k}(t) = \va_1$ in
the figure. In this case, no update is needed,
$\va_{k}(t+1)=\va_{k}(t)$, since $\va_{k}(t)$ satisfies both
conditions \eqref{constraint}, \eqref{normalization} and
$\|\va_{k}(t+1)-\va_{k}(t)\| = 0$. In the second case, we have
$\mathrm{sign} \left( \vec{c}(t+1)^T\va_{k}(t) \right)\neq
\mathrm{sign}\left(z_k(t+1)\right) $, e.g., $\va_{k}(t) = \va_2$ in
the figure. For this case, i.e., $\va_{k}(t) = \va_2$, we only need to
project $\va_{k}(t)$ to the half hyper sphere (shown as a half circle
in two dimensions in Fig.~\ref{projection}), which corresponds to the
constraints \eqref{constraint} and \eqref{normalization}. This
projection can be readily accomplished by first projecting
$\va_{k}(t)$ to the vector space perpendicular to $\vec{c}(t+1)$ and
then scaling the projected vector to have unit norm. This yields the
following
\[
\va_{k}(t+1) = \frac{\va_{k}(t)-\gamma(t+1)\frac{\vec{c}(t+1)^T\va_{k}(t)}{2\|\vec{c}(t+1)\|^2}\vec{c}(t+1)}
		         {\|\va_{k}(t)-\gamma(t+1)\frac{\vec{c}(t+1)^T\va_{k}(t)}{2\|\vec{c}(t+1)\|^2}\vec{c}(t+1)\|}
\]
update, where
\[\gamma(t+1) \defi 1 - \mathrm{sign}\left(z_k(t+1)\right) \mathrm{sign}\left(\vec{c}(t+1)^T\va_{k}(t)\right).\]

Here, \eqref{normalization} is needed to resolve the inherent
amplitude uncertainty in~\eqref{constraint} since the diffused sign
bit does not carry any amplitude information. We 
resolve the amplitude uncertainty in the final combination by
multiplying the unit norm estimate $\va_{k}(t+1)$ with the magnitude
of the local parameter estimate $\vphi_i(t+1)$. This scaling with the
norm of $\vphi_i(t+1)$ results in the rotated parameter estimation in
the direction of \(\va_{k}(t+1)\). After the construction of the
exchange estimates, the final estimate \(\vw_i(t+1)\) is calculated as
\begin{align*}
\vec{w}_i(t+1) = \lambda_{i,i} \vec{\phi}_i(t+1) + \|\vec{\phi}_i(t+1)\|\sum_{k \in \calN_i \setminus i} \lambda_{i,k} \va_{k}(t+1).
\end{align*}
\noindent
{\bf Remark 3:}
Fig.~\ref{projection} also demonstrates the update procedure for $\va_k(t)=\va_2$. The update is performed if the line, \(\vc_{\perp}\), perpendicular to \(\vc(t+1)\) passes through the shaded update region. Otherwise, the exchanged sign provides no new information and is discarded.

Alternatively, we can also resolve the amplitude uncertainty by using a sign LMS \cite{sayed_book} based approach. In this approach, at each node, we run an adaptive algorithm considering \(\vc(t+1)^T\vphi_k(t+1)\) as the desired data and \(\vc(t+1)\) as the regression vector. We then diffuse
 the sign of the error $\eps_{k}(t+1)\defi \vc(t+1)^T\vphi_k(t+1)-\vc(t+1)^T\va_{k}(t)$. Using the sign algorithm~\cite{sayed_book}, each node \(k\) can construct the exchange estimate as
\begin{align}
\va_{k}(t+1) = \va_{k}(t) + \sigma_{k}~\mathrm{sign}(\eps_{k}(t+1)) \vec{c}(t+1). \label{eq:sign}
\end{align}
Assuming \(\va_{k}(t)\)'s are initialized with the same values at each
node, \eqref{eq:sign} can be repeated at all neighboring nodes of
\(k\) to produce the same \(\va_{k}(t)\). In the next section, we
analyze the global stability of the algorithms in the mean.
\vspace{-0.1in}
\section{Stability Analysis}
We can write the reduced dimension diffusion~\eqref{eq:inv} and the
sign algorithm inspired diffusion~\eqref{eq:sign} approaches in a
compact form as
\begin{align}
&\vec{\phi}_i(t+1) = \vec{w}_i(t) + \mu_i \vu_i(t) e_i(t),\label{eq:update}\\
&\va_{k}(t+1) = \va_{k}(t) + \sigma_{k} \vc(t+1) h\left(\eps_{k}(t+1), \vc(t+1)\right)\label{eq:est}\\
&\vec{w}_i(t+1) = g_i\left(\vec{\phi}_i(t+1), \va_{k}(t+1);k \in \calN_{i}\setminus i\right),\label{eq:comb}
\end{align}
where \(\mu_i>0\) and $\sigma_{k}>0$ are the local learning rates, \(g_i(\cdot) \) is a combination function such as \eqref{comb1} and
\begin{align*}
& e_i(t) = d_i(t) - \vu_i(t)^T\vec{w}_i(t),\\
&\eps_{k}(t+1) = \vc(t+1)^T\left(\vphi_k(t+1)-\va_{k}(t)\right)
\end{align*} 
are the estimation and projected reconstruction errors. Here,
\(h\left(\eps_k(t+1),\vc(t+1)\right)\) is a generic function of
\(\eps_k(t+1)\) and  \(\vc(t+1)\), e.g., for the
scalar diffusion case \(h\left(\eps_k(t+1),\vc(t+1)\right)=
\left(\vc(t+1)^T\vc(t+1)\right)^{-1}\eps_k(t+1)\).

We define deviations from the parameter of interests as
\begin{align}
\Delta\vphi_k (t+1) &= \vec{w}_o - \vphi_k(t+1),\label{eq:dev_est}\\
\Delta\va_{k} (t+1) &= \vphi_k(t+1) - \va_{k}(t+1).\label{eq:dev_cons}
\end{align}
Substituting~\eqref{eq:dev_cons} into~\eqref{eq:comb}, we get the final estimate as
{\small \begin{align} 
\vec{w}_i(t+1) = \sum_{k \in \calN_i} \lambda_{i,k} \vphi_{k}(t+1) - \sum_{k \in \calN_i \setminus i} \lambda_{i,k}~\Delta \va_{k}(t+1).\label{comb}
\end{align}}
We then define the following global variables 
\begin{align*}
&\Delta\vphi(t) \defi \begin{bmatrix} \Delta\vphi_1(t) \\ \vdots \\ \Delta\vphi_N(t)\end{bmatrix},  \; \;
\Delta\va(t) \defi \begin{bmatrix} \Delta\va_{1}(t) \\ \vdots \\ \Delta\va_{N}(t)\end{bmatrix},\\
&\vec{U}(t) \defi \begin{bmatrix} \vu_1(t)&\ldots&\vec{0}\\ \vdots & \ddots & \vdots \\ \vec{0} & \ldots & \vu_{N}(t)\end{bmatrix}, \; \;
\vec{v}(t) \defi \begin{bmatrix} v_1(t) \\ \vdots \\ v_N(t) \end{bmatrix},
\end{align*}
where the vector dimensions are $(mN \times 1)$ and the matrix dimensions are $(mN \times N)$.

Using~\eqref{eq:update},~\eqref{eq:est},~\eqref{eq:dev_est},~\eqref{eq:dev_cons}, and~\eqref{comb}, we get 
\begin{align}
\Delta\vphi(t+1) =& \left(\vec{I} - \vec{D}\vec{U}(t)\vec{U}(t)^T\right)\vec{G}~\Delta\vphi(t) - \label{eq1} \\
		   &\left(\vec{I} - \vec{D}\vec{U}(t)\vec{U}(t)^T\right)\vec{\tilde{G}}~\Delta\va(t)+\vec{D}\vec{U}(t)\vec{v}(t), \nn
\end{align}
where \(\vec{G} \defi \Lambda \otimes \vec{I}_m\) is the transition matrix (and $\otimes$ is the Kronecker product), \(\vec{\tilde{G}} \defi \vec{G} - \mathrm{diag}\left(\vec{G}\right)\), \(\Lambda \defi \left[\lambda_{i,k} \right]\) is the combination matrix and \(\vec{D}\defi \mathrm{diag}\left([\mu_1,\mu_2,...,\mu_N]\right) \otimes \vec{I}_m\). 

We assume that the original parameter estimates \(\vphi_i(\cdot)\) vary slowly relative to the constructed estimates \(\va_i(\cdot)\) such that 
\begin{align*}
&\Delta\va_k(t) = \vphi_k(t)-\va_k(t) \cong \vphi_k(t+1)-\va_k(t) \mbox{ or}\\
&\Delta\va_k(t+1) = \vphi_k(t+1)-\va_k(t+1) \cong \vphi_k(t)-\va_k(t+1).
\end{align*}
Then the global update for the reconstructed parameters yields
\begin{align}
\Delta\va(t+1) = \left(\vec{I} - \vec{S}\vec{H}(t)\right)\Delta\va(t),\label{eq2}
\end{align}
where $\vec{S}\defi \mathrm{diag}\left([\sigma_1,\sigma_2,...,\sigma_N]\right) \otimes \vec{I}_m$ and $\vec{H}(t)$ is an appropriate transition matrix. As an example, for the scalar diffusion case we have 
\[\vec{H}(t) = \vec{I}_m \otimes \left(\frac{\vc(t+1)\vc(t+1)^T}{\vc(t+1)^T\vc(t+1)}\right).\]

For the single-bit diffusion,
$h(\eps_k(t+1),\vc(t+1))=\mathrm{sign}(\eps_k(t+1))$ is a nonlinear
function of $\eps_k(t+1)$, hence it is not straightforward to write
\eqref{eq2}. Although $h(\eps_k(t+1),\vc(t+1))$ is nonlinear, it can
be linearized using a Taylor series expansion. However, note that for
a sufficiently small step size $\sigma_k$ and Gaussian projection
vectors $\vc(\cdot)$, by the Price's theorem~\cite{sayed_book}, we can
write the expectation of the deviation $\Delta\va_k(t+1)$ as
\cite{sayed_book}
\begin{align*} {\small
\Delta\overline{\va}_k(t+1)=\Delta\overline{\va}_k(t)-\sigma_k\sqrt{\frac{2}{\pi}} \frac{E\left[\vc(t+1)\vc(t+1)^T\right]}{E\left[\eps_k^2(t+1)\right]}\Delta\overline{\va}_k(t).}
\end{align*}
Defining $\vec{F}(t+1)\defi \sqrt{\frac{\pi}{2}}~\mathrm{diag}\left([\eps_1^2(t+1),...,\eps_N^2(t+1)]\right)$ leads
\begin{align*} 
\overline{\vec{H}}(t)= \left[\overline{\vec{F}}(t+1) \otimes \vec{I}_m\right]^{-1}\left[\vec{I}_m \otimes \left(E\left[\vc(t+1)\vc(t+1)^T\right] \right)\right].
\end{align*}

Assuming the temporal independence of the projection signal
$\vc(\cdot)$ and the regression data $\vu_{k}(\cdot)$, taking the
expectation of both sides of~\eqref{eq1} and~\eqref{eq2}, letting
\(\vec{R} = E\left[\vec{U}(t)\vec{U}(t)^T\right]\) and assuming that
the variance $E\left[\eps_k^2(t+1)\right]$ terms do not 
depend on $\Delta\overline{\vphi}_k(t)$ terms, we get
\begin{align}
\begin{bmatrix} \Delta \overline{\vphi}(t+1) \\ \Delta\overline{\va}(t+1)\end{bmatrix} = 
\begin{bmatrix} \left(\vec{I}-\vec{D}\vec{R}\right)\vec{G} & \left(\vec{I}-\vec{D}\vec{R}\right)\vec{\tilde{G}} \\ \vec{0} & \vec{I}-\vec{S}\overline{\vec{H}}(t) \end{bmatrix}\begin{bmatrix} \Delta \overline{\vphi}(t) \\ \Delta \overline{\va}(t)\end{bmatrix},\label{dev}
\end{align}which covers both the reduced dimension and single bit diffusion
strategies. From \eqref{dev} we observe that our algorithms are stable
in the mean if
\(|\lambda\left(\vec{I}-\vec{S}\overline{\vec{H}}\right)| < 1\)
(provided that the full diffusion scheme is stable), where
$\lambda(\cdot)$'s are the eigenvalues. As example, for the scalar
case, assuming $\vec{c}(\cdot)$ are i.i.d. zero mean with unit
variance, then
\(|\lambda\left(\vec{I}-\vec{S}\overline{\vec{H}}\right)| < 1\) if and
only if $|1-\sigma_i|<1$ for all $i$.  Furthermore, the step sizes
$\sigma_i$ for the reconstruction algorithms could be chosen
accordingly for comparable convergence performance with the full
diffusion case. Following examples illustrate these results.
\vspace{-0.2in}
\section{Numerical Example and Concluding Remarks\label{sec:simulations}}
\begin{figure}[t!]
\centering
\includegraphics[width=1.5in]{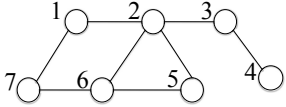}
\caption{The topology of the example network scenario with $N=7$ nodes.}
\label{scenario}
\end{figure}

In this section, we compare the introduced algorithms with the full
diffusion and no-cooperation schemes for the example network with
$N=7$ nodes given in  Fig.~\ref{scenario}. Here, we have stationary
data $d_i(t) = \vw_o^T\vu_i(t) + v_i(t)$ for $i = 1,2,...,N$, where
${\vu_i(t)}$ is i.i.d. zero mean with variance $0.1$,
$v_i(t)$ is i.i.d. zero mean with variance $0.1 \beta_i$, where
$\beta_i$ has uniform distribution $U[0,1]$, and $\vw_o \in
\mathbbm{R}^6$ is randomly chosen.

The combination matrix $\Lambda = [\lambda_{i,k}]$ is chosen as
\begin{align*}
\lambda_{i,k}=\left\{ \begin{array}{ll}
\frac{0.1}{\mathrm{max}(n_i,n_k)} & \mbox{if } i \neq k \mbox{ are linked}, \\ 
0 & \mbox{for $i$ and $k$ not linked},  \\
1-\sum_{k \in \calN_i \setminus i} \lambda_{i,k} & \mbox{for $i = k$},
\end{array}\right.
\end{align*}
where $n_i$ and $n_k$ denote the number of neighboring nodes for $i$
and $k$. Note that we modify the Metropolis rule such that
$\|\Lambda\| = 1$ and the exchanged information are weighted by $0.1$
to reduce variation while the exchange estimates $\vec{a}_k(\cdot)$'s are converging to the original parameter estimates $\vec{\phi}_k(\cdot)$'s.

The step sizes for the adaptation algorithms~\eqref{eq:update} of all
diffusion schemes are set to $\mu_i = 0.3$, $i = 1,2,...,N$. For
no-cooperation scheme, the step sizes are set to $\mu_i=0.03$. The
step sizes $\sigma_i$ for the reconstruction algorithms~\eqref{eq:est}
of the single-bit and the reduced dimension approaches are set as
$0.01$ and $0.5$, respectively. The randomized projection vectors
$\vc(t)$ are generated i.i.d. with standard deviation $0.1$ for the
single bit and $0.5$ for the reduced dimension diffusion strategies.
We point out that we set the learning rates for all algorithms such
that the final MSEs of all algorithms are the same for a fair
comparison.

\begin{figure}[t!]
\centering
\includegraphics[width=3.2in]{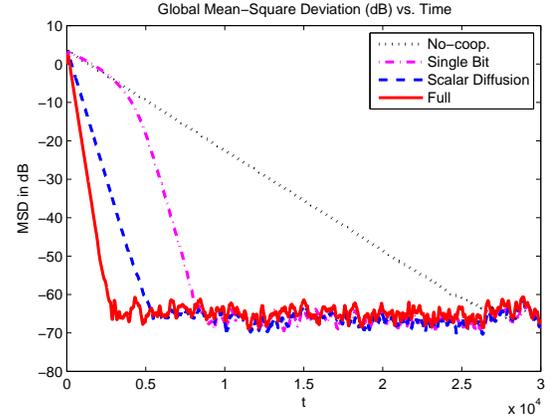}
\caption{Mean-square deviation (MSD) of various diffusion schemes.}
\label{plot}
\end{figure}

\begin{figure}[t!]
\centering
\includegraphics[width=3.2in]{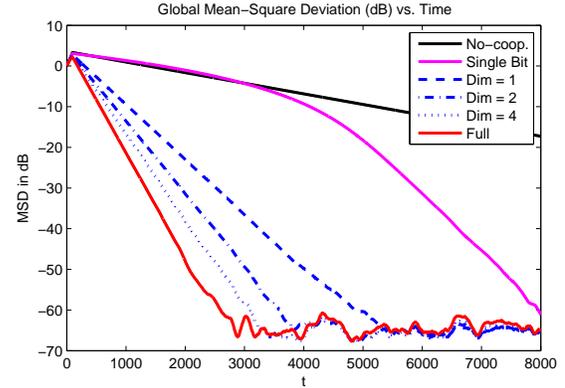}
\caption{Mean-square deviation (MSD) of reduced dimension diffusion strategy for different sizes.}
\label{plot2}
\end{figure}

In Fig.~\ref{plot}, we compare mean-square deviation of various
diffusion schemes in terms of their convergence performance for the
same steady state errors. As expected, in our simulations, the
introduced algorithms readily outperform the no-cooperation scheme in
terms of convergence performance. We observe from these simulations
that although we significantly reduce the amount of information
exchange, the introduced algorithms perform similar to the full
information case. To illustrate this further, in Fig.~\ref{plot2}, we
plot the performance of the reduced-dimension algorithm where we
gradually increase the number of dimensions that we kept. We observe
that as the number of dimensions increases, the reduced-dimension
algorithm gradually achieves the performance of the full information
case.

In this letter we introduce novel diffusion based distributed adaptive
estimation algorithms that significantly reduce the communication load
while providing comparable performance with the full information
exchange approaches in our simulations. We achieve this by exchanging
either a scalar or a single bit of information generated from random
projections of the estimated vectors at each node. Based on these
exchanged information, each node recalculates the estimates generated
by its neighboring nodes (which are then subsequently merged). We also
provide a mean stability analysis of the introduced approaches for
stationary data. This analysis can also be extended to mean-square and
tracking analysis under certain settings.


\vspace{-0.1in}
{\def\ninept{\def\baselinestretch{0.8}}
\ninept

\bibliographystyle{IEEEtran}
\bibliography{my_references}}
\end{document}